\documentclass[10pt,oneside]{article}

\usepackage[super,sort&compress,comma]{natbib} 
\usepackage[version=1]{mhchem}
\usepackage{times,mathptmx}
\usepackage{sectsty}
\usepackage{balance} 
\usepackage{url}

\usepackage{graphicx} 
\usepackage{lastpage}
\usepackage[format=plain,justification=raggedright,singlelinecheck=false,font=small,labelfont=bf,labelsep=space]{caption} 
\usepackage{amsmath}
\usepackage{color}
\usepackage{amssymb}
\usepackage{algorithmic}
\usepackage{setspace}
\usepackage{array}
\usepackage{float}
\usepackage{soul}
\usepackage{fancyhdr}
\title{An Efficient Descriptor Model for Designing Materials for Solar Cells}

\author{Fahhad H Alharbi $^{1,2}$, Sergey N Rashkeev $^{2}$, Fedwa El-Mellouhi $^{2}$, \\ Hans P L\"{u}thi $^{2}$, Nouar Tabet $^{1,2}$, and Sabre Kais $^{1,2}$ \\
1) College of Science and Engineering, Hamad Bin Khalifa University, Doha, Qatar \\
2) Qatar Environment and Energy research Institute (QEERI), \\
Hamad Bin Khalifa University, Doha, Qatar \\
falharbi@qf.org.qa}

\begin{document}
\maketitle

\begin{abstract}
\normalsize
An efficient descriptor model for fast screening of potential materials for solar cell applications is presented. It works for both excitonic and non-excitonic solar cells materials, and in addition to the energy gap it includes the absorption spectrum ($\alpha(E)$) of the material. The charge transport properties of the explored materials are modeled using the characteristic diffusion length ($L_{d}$) determined for the respective family of compounds. The presented model surpasses the widely used Scharber model developed for bulk-heterojunction solar cells [Scharber \textit{et al., Advanced Materials}, 2006, \textbf{18}, 789]. Using published experimental data, we show that the presented model is more accurate in predicting the achievable efficiencies. Although the focus of this work is on organic photovoltaics (OPV), for which the original Scharber model was developed, the model presented here is applicable also to other solar cell technologies. To model both excitonic and non-excitonic systems, two different sets of parameters are used to account for the different modes of operation. The analysis of the presented descriptor model clearly shows the benefit of including $\alpha(E)$ and $L_{d}$ in view of improved screening results.
\end{abstract}

\section{Introduction}
There has been a remarkable thrust toward developing cost-effective photovoltaics (PV) in the past two decades \cite{A02,H01,W02,FH1,FH2}. Different materials and device concepts have been deployed and the highest achieved conversion efficiency so far is 44.7\% by quadruple junction using III-V materials \cite{D01}. As for the market, it is dominated by the conventional \ce{Si} solar cells. Nonetheless, dye sensitized solar cells (DSSC), organic photovoltaics (OPV), and the recently emerged hybrid perovskite solar cells could become more cost effective and competitive if produced at large scale \cite{S01}.

In principle, an efficient single-junction solar cell can be made of any semiconducting material with an energy gap ($E_g$) ranging between 1.0 and 1.7 eV and with reasonable transport to allow the photo-generated carrier to be collected \cite{N01,M01,K01,FH1,FH3}. Hence, many organic and inorganic semiconductors have been used to make solar cells \cite{FH1,FH3}. The selection was mostly based on known materials as, till recently, experimental data was the main source for screening materials for solar cells. Despite the rich data, this certainly limits the screening space. However, the sophisticated computational capabilities have provided an alternative route to explore new materials for solar cells much beyond the rich experimental data. There are many initiatives in this regard. Among the most noticeable ones is the Clean Energy Project at Harvard University \cite{H02,O01,H07}. It is a high-throughput discovery and design program for the next generation of OPV materials. By 2013, 2.3 million of organic molecules and polymers were analyzed using more than 150 million density functional theory calculations to assess their applicability for solar cells \cite{H07}.

The Clean Energy Project, like other initiatives \cite{L03,C06,C07}, is usually based on atomistic calculations, which are then fed into empirical descriptor models to assess the potential of the studied material for photovoltaics. The commonly used descriptor model, at least within the OPV community, is the one proposed by Scharber \cite{SM1}, a one-parameter model based on the computed $E_g$, in which the open circuit voltage ($V_{oc}$) is assumed to be a fixed reduction of $E_g$ defined (by the Scharber model) as the difference between the highest occupied molecular orbital (HOMO) of the donor and the lowest unoccupied molecular orbital (LUMO) of the acceptor. The short circuit current ($J_{sc}$) is estimated as a fraction of the current resulting from absorbing all incident photons above $E_g$ while the fill factor $FF$ is set to a fixed value. Usually, the $FF$ and the scaling parameter for $J_{sc}$ are both set to a value of 0.65. Although these approximations for $V_{oc}$ and $FF$ appear reasonable, the assumption that all the photons above $E_g$ are absorbed and a fraction of them is extracted as current is an extreme oversimplification. These assumptions ignore the inhomogeneity of the absorption spectrum. Furthermore, it assumes that the transport is highly efficient and that the diffusion length is much larger than the absorption length so that the detailed-balance fraction of the photogenerated carriers can be collected. Practically, it is important to consider in more detail the absorption spectrum and the transport limitations.

In this paper, we propose a descriptor model where the absorption spectrum ($\alpha(E)$) is obtained from the same electronic structure calculations used to determine $E_g$. $E$ is the photon energy in eV, which is used as the unit for energy throughout this paper. These are the only atomistic calculations needed here. In addition, the transport is characterized by the diffusion length ($L_{d}$), which is a measure for the mean distance that an excited carrier can cross through random diffusion before recombining. Calculating $L_{d}$ needs lengthy calculations, which would make combinatorial screening prohibitively expensive. To avoid this, each material is given a value for $L_d$, which is characteristic for the family of compounds it belongs to.

The focus in this paper is on OPV as in the Scharber model. Nonetheless, the same model is applicable to other PV technologies. Yet, due to the slightly different operations between excitonic (such as OPV) and non-excitonic (e.g., inorganic semiconductor cells) solar cells, two distinct sets of parameters should be used. For non-excitonic solar cells, the binding energy of excitons is small and hence the exciton can be dissociated thermally or by potential gradient. On the other hand, in excitonic cells, where the binding energy is large, the heterojunction band offset is needed to dissociate excitons. Thus, a considerable additional loss in the voltage is unavoidable. Therefore, it is essential to make distinction between these two classes of cells.

We intend to apply our model for large scale virtual screening of organic compounds, where the absorption spectra fluctuate considerably. Hence, we expect that by taking into account the details of their absorption spectrum, we will be able to better discriminate between candidate compounds. The initial validation analysis clearly shows the merit of including $\alpha(E)$ and $L_{d}$ in the descriptor model. Just as an example, the Scharber model suggests that copper phthalocyanine (CuPc) is better than the parent squaraine (SQ) donor. However, by including $\alpha(E)$, the improved model shows that SQ should be more efficient than CuPc if the film thickness is less than 100 nm, which is within the normal range of OPV donor thickness \cite{H03,F03}. 

\section{The proposed descriptor model}

As known, PV efficiency ($\eta$) is commonly expressed as
\begin{equation}
\label{SCGenEff}
	\eta = \frac{V_{oc} \, J_{sc} \, FF}{P_{in}}
\end{equation}
where $P_{in}$ is the input power density. Certainly, many factors contribute to $V_{oc}$, $J_{sc}$, and $FF$. The main factors are materials related. Yet, the device design, fabrication quality, and operational conditions play major roles as well. For materials screening, it is reasonable to assume that the device design and quality are optimized. So, the merit of the material's potential for photovoltaic depends mainly on its optoelectronic properties. In this section, relations are proposed to link the materials properties to practical estimations for the maximum obtainable values for $V_{oc}$, $J_{sc}$, and $FF$ and hence the efficiency.

For OPV technology, the most efficient cells are bulk heterojunction devices (BHJ). Conceptually, for materials screening, BHJ device requires multi- purpose multi- dimensional screening; i.e. a matrix of possible devices need to be screened based on a set of acceptors and a set of donors that fulfill the requirements to make a working solar cell. If the sets are small, then, the two-dimensional screening is possible. Otherwise, it can become intractable. So, most of the related large-scale screening is performed for single-layer OPV \cite{SM1,H02,O01}. However, since BHJ devices allow for greater thickness than single layer ones, the assumed thickness should be larger than the actual ``exciton'' diffusion length ($L_{Xd}$), which is usually less than 100 nm \cite{H03,F03,V01}. In this work, we follow the same track of materials screening for single layer OPV, but assume that the thickness is equal to the nominal average of BHJ devices, i.e. around 150 nm \cite{H03,F03}. For practical reasons, the focus is on finding a small set of promising donor materials, for which it will be later possible to find matching acceptors.

The proposed descriptor model parameters will be based on the best experimentally reported efficiencies for different organic, inorganic, and organometallic materials. These data are tabulated in Appendix A (Tables \ref{InorgHmJ}, \ref{InrgHtJ}, and \ref{OrgHtJ}). There are some better reported efficiencies; unfortunately, there are no details about the performance parameters of these cells. So, we limit the analysis to the best reported cells with full details. The used reference materials are:
\begin{description}
\item[Excitonics:] SQ (2,4-bis[4-(N,N-diisobutylamino)-2,6-dihydroxyphenyl]squaraine), DTS (5,5-bis(4-(7-hexylthiophen-2-yl)thiophen- 2-yl)-[1,2,5]thiadiazolo[3,4-c]pyridine-3,3-di-2-ethylhexylsilylene-2,2'-bithiophene), CuPc (copper phthalocyanine), ZnPc (zinc phthalocyanine), DBP ((dibenzo([f,f']-4,4',7,7'-tetraphenyl)diindeno[1,2,3-cd:1',2',3'-Im]perylene), P3HT (poly(3-hexylthiophene-2,5-diyl), and PTB7 (poly[ [4,8-bis[(2-ethylhexyl)oxy]benzo[1,2-b:4,5-b0]dithiophene-2,6-diyl][3-fluoro-2-[(2-ethylhexyl)carbonyl]thieno[3,4–b]thiophenediyl]]),
\item[Non-excitonics:] \ce{Si}, \ce{GaAs}, \ce{InP}, \ce{GaInP}, \ce{CdTe}, \ce{CuIn$_x$Ga}$_{(1-x)}$\ce{Se2} (CIGS), and \ce{(CH3NH3)PbI3} (\ce{MAPbI3}).
\end{description}

\begin{figure} [pt]
\centering
\includegraphics[width=2.4in]{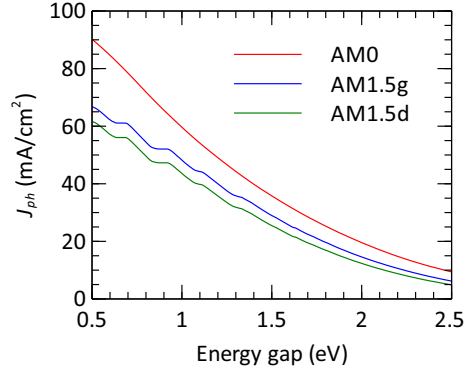}
\caption{$J_{ph}$ vs. $E_g$ corresponding to AM0, AM1.5g, and AM1.5d spectra.}
\label{JphEx}
\end{figure}

Before presenting the descriptor model, it is useful to discuss the solar photon flux density ($\phi_{ph}$) and to introduce simple approximations for the maximum obtainable current density ($J_{ph}$). The reference density is tabulated by the American Society for Testing and Materials standard (ASTM G173-03) for AM0, AM1.5g, and AM1.5d \cite{A03}. For flat-panels, AM1.5g is more appropriate and it will be used in this paper. If all the photons above a given $E_g$ are absorbed (i.e., the reflection is neglected and the thickness of the absorbing material is large enough) and each photon were to generate one exciton, the maximal photo-generated current $J_{ph}$ is
\begin{equation}
	\label{Jph1}
	J_{ph} = q \int_{Eg}^\infty \phi_{ph}(E) \, dE
\end{equation}
where $q$ is the electron charge. $J_{ph}$ for the different solar spectra are shown in Fig. \ref{JphEx}. In the Scharber model, $J_{sc}$ is assumed to be a fraction of $J_{ph}$ associated with AM1.5g spectrum. 

As $J_{ph}$ is used routinely in solar cells calculations, it would be useful to approximate $J_{ph}$ as a function of $E_g$ in the target range between 1 and 2 eV. This will be used later to develop the improved model. Using the data shown in Fig. \ref{JphEx}, three possible expressions for approximating $J_{ph}$ as a function of $E_g$ can be suggested:
\begin{equation}
	\label{JphA1}
	\tilde{J}_{ph,1} = a_1 \exp \left( -b_1 E_g \right),
\end{equation}
\begin{equation}
	\label{JphA2}
	\tilde{J}_{ph,2} = A - \delta J E_g + a_2 \exp \left( -b_2 E_g \right),
\end{equation} 
and
\begin{equation}
	\label{JphA3}
	\tilde{J}_{ph,3} = a_3 \exp \left( -b_3 E_g^{c_3} \right).
\end{equation} 

\begin{figure} [pt]
\centering
\includegraphics[width=2.4in]{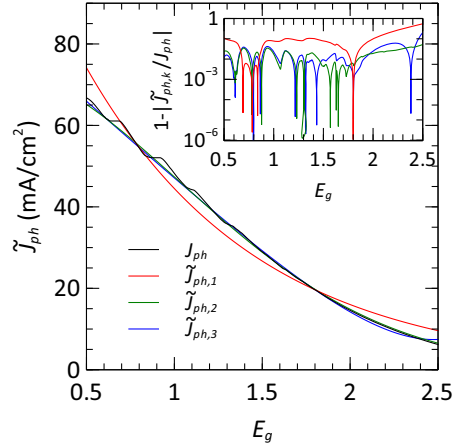}
\caption{The approximated $J_{ph}$ vs. $E_g$. In the inset, the errors are plotted vs. $E_g$ in logarithmic scale.}
\label{JphApp}
\end{figure} 

\noindent The parameters resulting in the best fit (Fig. \ref{JphApp}) are:
\begin{itemize}
\item for  $\tilde{J}_{ph,1}$, $a_1=123.62$ and $b_1=1.0219$,
\item for  $\tilde{J}_{ph,2}$, $a_2=0.09097$, $b_2=-2.14$, $A=85.02$, and $\delta J=38.69$,
\item for  $\tilde{J}_{ph,3}$, $a_3=73.531$, $b_3=0.440$, and $c_3=1.8617$.
\end{itemize}
For its simplicity and good accuracy, the expression for $\tilde{J}_{ph,3}$ will be used in this work. It will be referred to as $\tilde{J}_{ph}$, and, correspondingly, the numerical index will be dropped also for the fitting parameters $a$, $b$, and $c$.

Parameters $V_{oc}$, $J_{sc}$, and $FF$ are tightly coupled. Thus, to have estimations using only $E_g$, $\alpha(E)$, and $L_d$, many approximations are needed. In this work, we will start by estimating $V_{oc}$ as a function of $E_g$, where the extracted current is assumed as a fraction of $J_{ph}$. Then, $E_g$, $\alpha(E)$, and $L_d$ will be used to estimate $J_{sc}$. Finally, $FF$ is estimated based on $V_{oc}$.

\subsection{The open-circuit voltage ($V_{oc}$)}

Theoretically, $V_{oc}$ is the maximum voltage that a solar cell can apply to an external load. It is essentially the difference between electron and hole quasi-Fermi levels resulted from photo-excitation. Typically, it is assumed to be upper bounded by $E_g/q$, which is standardly defined -unlike the Scharber model- as the difference between HOMO and LUMO of a single absorbing materials. In the highly unlikely case of extreme charge accumulation, it can exceed the gap. Many relations between $V_{oc}$ and $E_g$ were suggested \cite{K01,A01,M01,W01,N01,B01}. Almost all of them are based on the Shockley diode equation (assumed ideal with the identity number set to unity) when the net current vanish. This leads to
\begin{equation}
	\label{ScDiode}
	V_{oc} = \dfrac{k_B T}{q} \ln \left( \dfrac{J_{sc}}{J_0} + 1 \right)
\end{equation}
where $k_B$ is the Boltzmann constant, $T$ is the cell temperature, and $J_0$ is the reverse saturation current density. The differences between the suggested models are due to the different assumptions for $J_{sc}$ and $J_0$. For $J_{sc}$, in this subsection, it is assumed to be a fraction of $J_{ph}$. This is acceptable as the scaling constant will be considered by the fitting parameters. As for $J_0$, many models and empirical equations were suggested \cite{F01,N02,W03,C01}. Among the best approximations is the Wanlass equation \cite{W03}, where the values of $J_0$ of many of the commonly used semiconductors are fitted to very high accuracy. According to his model,
\begin{equation}
	\label{Wanlass}
	J_0 = \beta (E_g) T^3 \, \exp \left( - \dfrac{E_g}{k_B T} \right)
\end{equation}
where
\begin{equation}
	\label{BetaWanlass}
	\beta(E_g) = 0.3165 \, \exp \left( 2.192 E_g \right).
\end{equation}
in $mA/cm^2K^3$. Theoretically, $\beta$ should be constant. However, $E_g$ dependence is introduced empirically as a correction for homojunction ``solar cells'' operation \cite{W03,C01,C05}. The same form was suggested also for OPVs, which are heterojunction devices, but with slightly smaller value for the prefactor \cite{P02,K07}. So, by applying a fraction of $\tilde{J}_{ph}$ (Eq. \ref{JphA3}) and $J_0$ in Eq. \ref{ScDiode} and by considering the fact that $J_0$ is very small quantity, we obtain
\begin{equation}
	\label{VocA}
	\tilde{V}_{oc}=E_g-\tilde{V}_L
\end{equation}
where at room temperature (at 300 K)
\begin{equation}
	\label{VocA1}
	\tilde{V}_L=0.0114 E_g^{1.8617} + 0.057 E_g + V_{L0}
\end{equation}
and $V_{L0}$ is used as a fitting parameter (thus the differences in the prefactors of $J_0$ are accounted for).

\begin{figure}[t]
\centering
\includegraphics[width=2.4in]{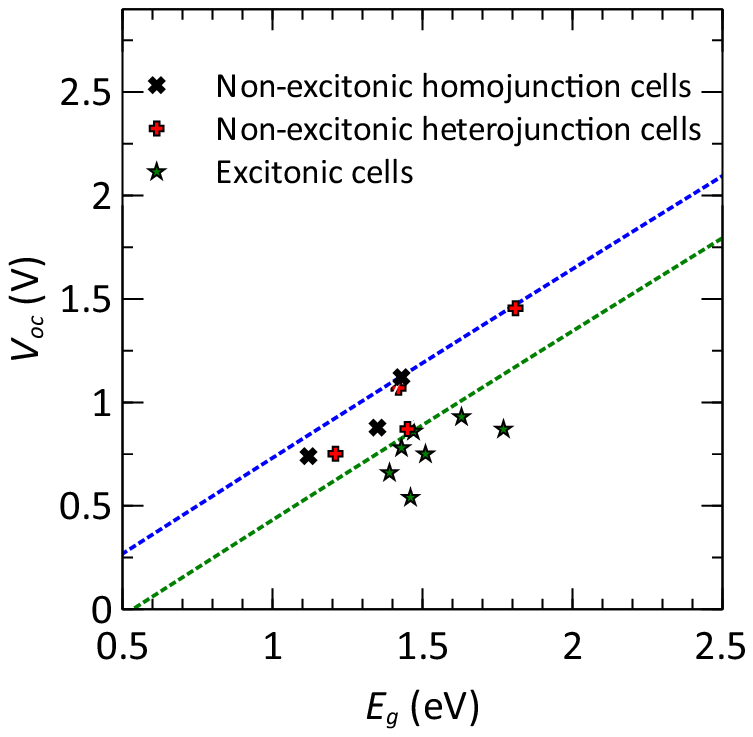}
\caption{The obtained $V_{oc}$ vs. $E_g$ for the reference materials. The black marks are for inorganic homojunction cells (non-excitonic), the red marks are for the inorganic and organometallic heterojunction junction solar cells (non-excitonic), and the green ones are for the organic cells (excitonic). The two dotted lines are for different values of $V_L$. The blue line is for $V_{L0}=0.2$ V, which fits non-excitonic cells, and the green line is for $V_{L0}=0.5$ V for excitonic cells.}
\label{VocC}
\end{figure}

In homojunction solar cells, the losses are mainly due to the materials and the excitation; i.e. the losses due to the device design are -in principle- avoidable. For heterojunction devices, the energy offsets between the layers add to the voltage loss. For non-excitonic solar cells, the binding energy of exciton is small and hence it can be dissociated thermally or by a potential gradient. As a result, the heterojunction offset can be made small. On the other hand, in excitonic cells, the binding energy is large and the band offset is used to dissociate excitons. Thus, a considerable additional loss in the voltage is unavoidable. Therefore and as aforementioned, it is essential to make distinction between the two classes of excitonic and non-excitonic cells. The original Scharber model considers the reduction due to band offset. However, this is routinely ignored in materials screening as it adds extra constraint on the acceptor. 

Fig. \ref{VocC} maps the obtained $V_{oc}$ to $E_g$ for the reference materials. Clearly for non-excitonic cells, the blue line ($V_{L0}=0.2$ V) line provides a good estimation for the upper limit. On the other hand, the maximum experimentally measured $V_{oc}$ values for excitonic cells are much lower. This is mainly due to the sizable, yet needed heterojunction band offset. Thus, larger $V_{L0}$ is indispensable. Here, this parameter is set to the lowest reported voltage loss in organic cells. As can be seen in Table \ref{OrgHtJ} (in the appendix), it is 0.61 V for SQ based solar cell leading to $V_{L0}\thickapprox 0.5$ V.

\subsection{The short-circuit current ($J_{sc}$)}

As stated in the introduction, in the Scharber model, $J_{sc}$ is assumed to amount to a constant fraction of $J_{ph}$. Usually, 0.65 is used as the scaling parameter. So,
\begin{equation}
	\label{JscSch}
	\tilde{J}_{sc,Sch} = 0.65 \, J_{ph}
\end{equation}
The two most crucial deficiencies of the Scharber model, namely the assumption of a homogeneous absorption spectrum (above the band gap) and that the transport is very efficient such that $L_d$ is much larger than the absorption length, can be addressed by explicitly considering the spectral inhomogeneity (using $\alpha(E)$) and by introducing a proper characterization of $L_d$. This shall result in improved predictions while not over-complicating the descriptor model.

The absorption spectrum $\alpha(E)$ can be computed by means of electronic structure calculations, often based on semiclassical approaches. These calculations also provide numerical values for $E_g$. So, there is no additional atomistic calculations needed for $\alpha(E)$. However, the calculation of $\alpha(E)$ from these common inputs can be computationally expensive. This fact shall be considered during the design of high-throughput screening.

Commonly, $\alpha(E)$ is calculated by semiclassical approach where the electrons are treated quantum mechanically through the electronic structure and the field is treated classically. The details vary based on the used method for electronic structure calculations. From electronic structure calculations, the complex dielectric function ($\varepsilon(E) = \varepsilon_1(E) + i \varepsilon_2(E)$) can be calculated. $\varepsilon_2(E)$ is calculated by considering all the possible transitions from occupied to unoccupied states. For each transition, its contribution into $\varepsilon_2(E)$ is proportional to the square of the matrix element. Then, $\varepsilon_1(E)$ is calculated from $\varepsilon_2(E)$ using the Kramers-Kronig transformation. Finally, both the refractive index ($n(E)$) and $\alpha(E)$ are calculated from the relation $\sqrt{\varepsilon_1(E) + i \varepsilon_2(E)} = n(E) + i \alpha(E) \, \hbar \, c / q \, E$ where $\hbar$ is the Planck constant and $c$ is the speed of light.

As aforementioned, the transport is commonly characterized by $L_{d}$. Calculating $L_{d}$ requires very time-consuming computing, which will complicate the materials screening process. To avoid that, each material is given a value of $L_{d}$, which is characteristic for the family of compounds it belongs to. 

Many parameters govern $L_d$. Some of them are related to the intrinsic properties of materials and many others are due to the fabrication quality. For non-excitonic cells, the minority carrier diffusion is the main process and it is limited mainly by material growth quality. For defect free indirect band-gap materials, lifetimes are in milliseconds and the mobilities are high which give rise of few hundred microns to $L_d$ \cite{S02,F02}. However, for direct gap materials, the lifetime is significantly reduced because of the band-to-band recombination. Thus $L_d$ is reduced to a range between few microns and few tens of microns \cite{L01,M02,G01}. As for organometallic materials, $L_d$ is estimated to be more than 1  $\mu$m for methylammonium lead iodide \cite{S03}. 

\begin{figure*}[t]
\centering
\includegraphics[width=5in]{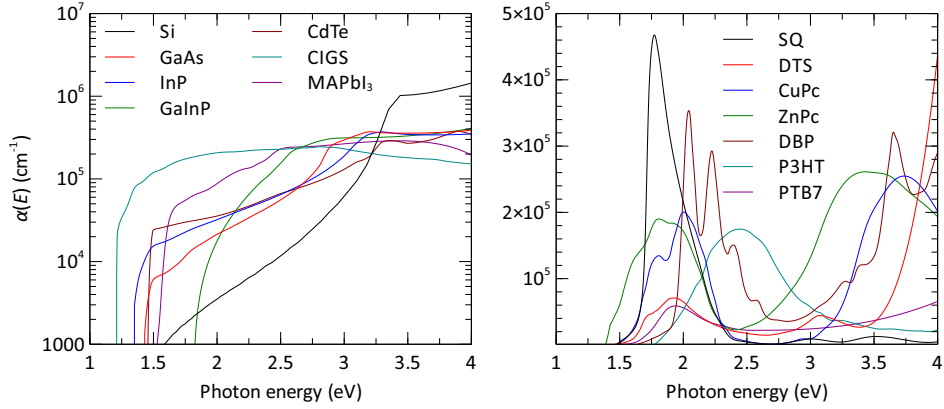}
\caption{The absorption coefficients of the reference materials; left: non-excitonic cells, right: excitonic cells. The data are extracted from various sources \cite{S04,W04,T01,G04,W05,D02,F04,S05}.}
\label{AbsSp}
\end{figure*}

On the other hand, the main limiting factor in excitonic solar cells is the exciton diffusion length \cite{G02,G03}. The exciton cannot dissociate at the same location at which it was generated. Rather, it has to travel by hopping to the nearest interface to dissociate. Thus, the transport is limited by exciton diffusion rather than the free carrier diffusion. Exciton diffusion length in organic solar cell materials is normally less than 0.1 $\mu$m \cite{H03,F03,V01}. In this work, we assume the following values for $L_d$ as characteristic for the following material families:
\begin{itemize}
\item for indirect-gap semiconductors, $L_d \thickapprox 200\mu$m,
\item for direct-gap semiconductors, $L_d \thickapprox 10\mu$m,
\item for organometallic semiconductors, $L_d \thickapprox 0.6\mu$m,
\item for excitonic cells, $L_d \thickapprox 0.1\mu$m.
\end{itemize}

Conceptually, $J_{sc}$ is the difference between photo-generated and recombination currents, i.e.
\begin{equation}
	\label{Jsc1}
	J_{sc} = J_g - J_r
\end{equation}
In an ideal situation, it is assumed that the thickness of the absorber layer is so large that all photons above $E_g$ are absorbed. Practically, the carrier collection and hence the absorber layer thickness are limited by $L_d$. Therefore, the maximum possible photo-generated current is \cite{P03,K02,K03,M04}
\begin{equation}
	\label{Jsc2}
	J_g = q \int_{Eg}^\infty \int_{-\pi/2}^{\pi/2} \phi_{ph}(E) \, P \left( \theta , \theta_{inc}, E \right) \\ \left[ 1-e^{-\alpha(E)L_d/\cos(\theta)} \right] \, d\theta \, dE
\end{equation}
where $P \left( \theta , \theta_{inc}, E \right)$ is an angular distribution function that accounts for the scattering of the light at angle $\theta$ in the absorbing layer depending on the incidence angle $\theta_{inc}$ and photon energy. The scattering results in increasing -positively- the optical path of the light in the absorbing layer by a factor of $1/\cos(\theta)$. From a device-performance perspective, it is important to have $L_d$ much larger than the absorption length ($L_{\alpha} \propto 1/\alpha(E)$). If $L_d \gg L_{\alpha}$, the second term in the square bracket gets diminished and $J_g$ increases. Otherwise,   $J_g$ is reduced considerably.

\begin{figure*} [ht]
\centering
\includegraphics[width=5in]{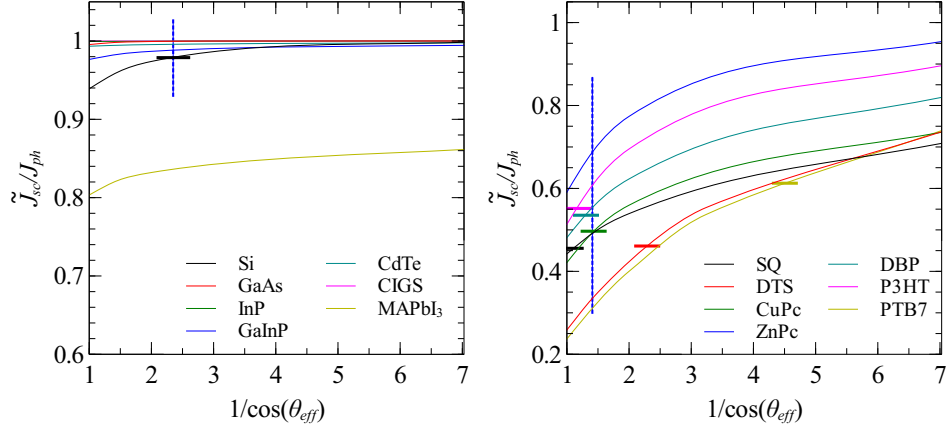}
\caption{The ratio between $\tilde{J}_{sc}$ and  $J_{ph}$ for the reference materials at different $1/\cos(\theta)$; left: non-excitonic cells, right: excitonic cells. The small marks correspond to the maximum reported ratios. For the materials without shown marks, the ratios match the maximum possible values without scattering. This is mainly due to their strong absorptions and very efficient transports.}
\label{AbsSpTh}
\end{figure*}

Obviously, $P \left( \theta , \theta_{inc}, E \right)$ depends on many factors such mainly related to the films morphology and microstructure and the structure of interfaces. For the modeling of the distribution function, many different distributions were suggested \cite{B02,C02,K02,K03,C03,K04}. In this work, to keep the model simple, we can combine the scattering effects in a single effective angle $\theta_{eff}$. So, $J_g$ becomes
\begin{equation}
	\label{Jsc2e}
	\tilde{J}_g \left( E, \alpha(E), \theta_{eff} \right) = q \int_{Eg}^\infty \phi_{ph}(E) \, \\ \left[ 1-e^{-\alpha(E)L_d/\cos(\theta_{eff})} \right] \, dE
\end{equation}
The way $\theta_{eff}$ is determined will be shown at the end of this subsection.

As for the recombination, there are many mechanisms contributing to it. This is accounted for empirically by $L_d$ and it can be adjusted further by a proper fitting of $\theta_{eff}$. So, $J_{sc}$ can be approximated by a similar form of Eq.-\ref{Jsc2e}, but with a slightly different $\theta_{eff}$, that will be determined based on the actual performances and absorption spectra of the reference materials.

Fig. \ref{AbsSp} shows the absorption spectra of the reference materials, which are extracted from various sources \cite{S04,W04,T01,G04,W05,D02,F04,S05}. For the known non-excitonic cells, it is evident that due to the extended absorption spectrum above $E_g$, $\alpha(E)$ is smooth, whereas the organic materials show a strongly fluctuating bands. For example, the absorption of SQ is strong only between 1.5 and 2.3 eV. Thus, a large portion of solar radiation is not absorbed due to the fact that the device thickness is small.

To determine $\theta_{eff}$, the ratios  $\tilde{J}_{sc} / \tilde{J}_{ph}$ are calculated and plotted against $1/\cos(\theta_{eff})$ (Fig. \ref{AbsSpTh}). For non-excitonic cells, the effect of $\theta_{eff}$ for most materials is negligible due to their strong absorption and due to the fact that the growth quality of the studied materials are high and hence the assumed $L_d$ is large. The exceptions are for Si due to its weak absorption and for \ce{MAPbI3} due to its relatively short $L_d$. For Si, $\theta_{eff} \thickapprox \pi / 2.75$ is needed to match the obtained $J_{sc}$. So, this value will be used for non-excitonic cells. For excitonic cells, $\theta_{eff} = \pi / 4$ is a good approximation for most of the studied materials. The exception is for PTB7, where the difference between the reported $J_{sc}$ and the calculated value from absorption spectrum is high. This can be due to an extremely efficient light trapping used to make the cell \cite{H05}. However, to match most of the reported maximum values, $\theta_{eff} = \pi / 4$ is suitable and will be used for excitonic cells.

\subsection{The fill factor ($FF$)}

The third performance parameter is the fill factor ($FF$). Practically, many physical mechanisms contribute to it and consequently many models have been suggested to estimate it \cite{Q01,T02,G05,G06}. Generally, the suggested models are based on the relationship between current and voltage; but with different assumptions on the causes and values of shunt and series resistances. $FF$ is usually represented as a function of $V_{oc}$, which depends as shown above on $E_g$.

\begin{figure} [pt]
\centering
\includegraphics[width=2.3in]{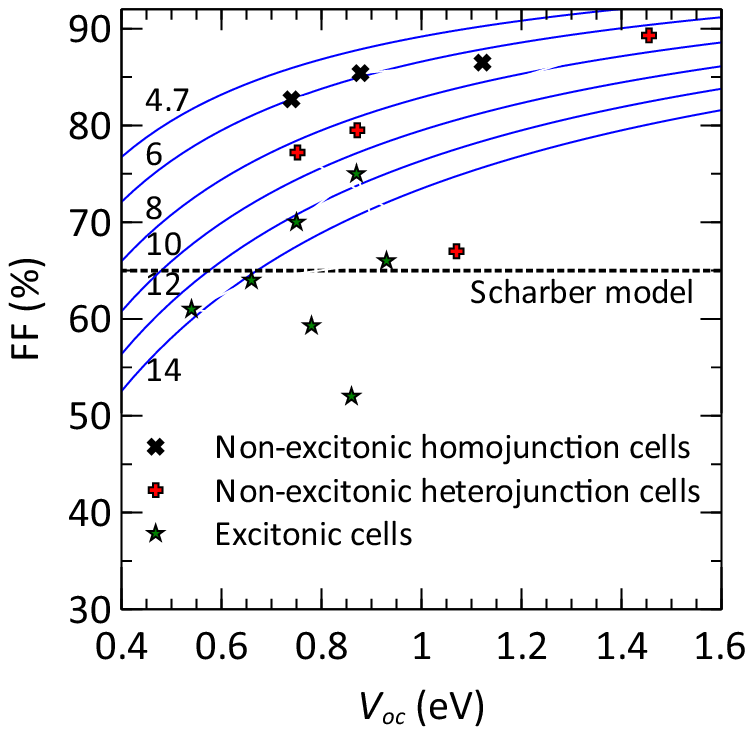}
\caption{The obtained $FF$ vs.$V_{oc}$ for the reference materials. The black marks are for inorganic homojunction cells (non-excitonic), the red marks are for the inorganic and hybrid heterojunction solar cells (non-excitonic), and the green ones are for the organic cells (excitonic). The solid lines are for 6 different values of $a$ while the dotted black line is the value suggested by Scharber model (Table \ref{ComparisonTable}).}
\label{FFFig}
\end{figure}

One of the simplest -yet reasonably accurate- forms suggested by Green \cite{G06} for conventional inorganic semiconductors is 
\begin{equation}
	\label{FF01}
	FF = \frac{V_{oc}}{V_{oc}+a \, k_B T}
\end{equation}
Originally, he suggested $a=4.7$. However, this factor can be adjusted for other solar cell technologies. Based on the best reported cells, $a=6$ and $a=12$ fit better the upper limits of the measured $FF$ for non-excitonic and excitonic solar cells respectively as shown in Fig. \ref{FFFig} where $T$ is the room temperature. 

\section{Model Implementation and Evaluation}

\begin{table*}[pt]
\small
\centering
\renewcommand{\arraystretch}{1.5}
\caption{A summary of the original Scharber model and proposed model for both excitonic and non-excitonic solar cells. For the original Scharber model, $\Delta{V}$ is the band off-set.}
\begin{tabular}{|c|c|c|c|}
\hline 
 & The original model & The proposed model for excitonic cells & The proposed model for non-excitonic cells \\ 
\hline 
$V_{oc}$ & $E_g -0.3 - \Delta{V} $ & $E_g - 0.5 - 0.0114 E_g^{1.8617} -0.057 E_g$ & $E_g - 0.2 - 0.0114 E_g^{1.8617} -0.057 E_g$ \\ 
$J_{sc}$ & $0.65 \, J_{ph}(E_g)$ & $\tilde{J}_g \left( E_g, \alpha(E), \pi/4 \right)$ & $\tilde{J}_g \left( E_g, \alpha(E), \pi/2.75 \right)$ \\  
$FF$ & 0.65 & $V_{oc} / ( V_{oc} + 12 k_B T_c ) $ & $V_{oc} / ( V_{oc} + 6 k_B T_c )$ \\ 
\hline 
\end{tabular} 
\label{ComparisonTable}
\end{table*}

\begin{figure}[hp]
\centering
\includegraphics[width=2.3in]{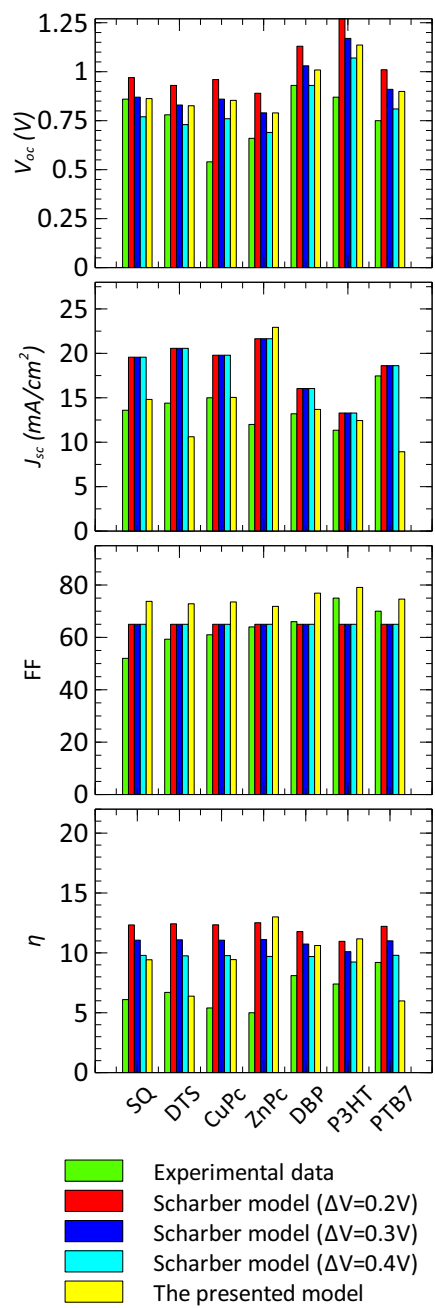}
\caption{Comparison for organic (excitonic) solar cell performances of the relevant reference materials as estimated by the presented and the Scharber (for 3 different values for $\Delta{V}$) models, and as experimentally published for the best reported cells.}
\label{Comp1}
\end{figure}

The parameterized expressions used for the three performance factors ($V_{oc}$, $J_{sc}$, and $FF$) in the original Scharber model and the presented descriptor are summarized in Table \ref{ComparisonTable}. 

In the first analysis, we compare the predictions of the two models for excitonic materials with the available experimental data (Table \ref{OrgHtJ}). For the original Scharber model, three reasonable values \cite{D03,H02,H07} for $\Delta{V}$ are assumed in the analysis; namely 0.2, 0.3, and 0.4 V. The results are shown in Fig. \ref{Comp1}. As can be observed, the improved model outperforms considerably the original Scharber model in the estimations of $J_{sc}$, except for PTB7. For $V_{oc}$, the presented model provides good estimation for most of the studied materials. As for the original Scharber model, this depends obviously in $\Delta{V}$. For $\eta$, the presented model outperforms in most cases the original one. The original model performs generally better only in the estimation of $FF$. However, $FF$ depends extremely on the device design and optimization and unlike other parameters which are mostly materials dependent. The improved model suggests that the obtained $FF$ values are smaller than the predictions. Thus, there is a reasonable room for improvement through device optimization.

\begin{figure}[hp]
\centering
\includegraphics[width=2.3in]{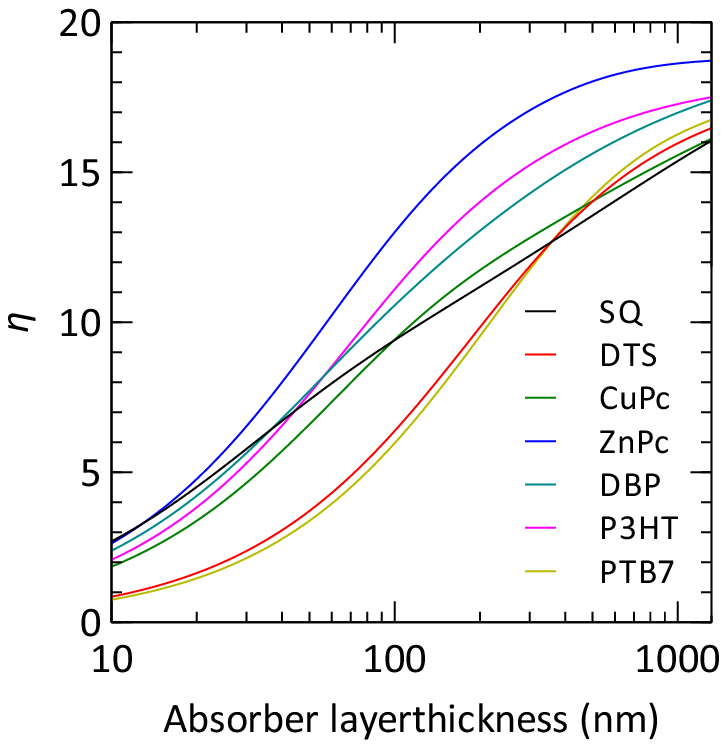}
\caption{The estimated solar cell efficiencies for the reference organic (excitonic) materials as a function of the absorbing layer thickness.}
\label{Comp2}
\end{figure}

To illustrate the importance of considering $L_d$, the presented method is used to estimate conversion efficiencies of the array of organic materials studied as a function of thicknesss as shown in Fig. \ref{Comp2}. Clearly, material potential ranking varies with the thickness. For example, for very thin films ($L < 50 $ nm), SQ is predicted to show better efficiency when compared to all other materials, except ZnPc. Also, PTB7 and DTS based solar cells would result in better efficiencies when compared to CuPc and SQ only for relatively thicker films.

\begin{figure}[hp]
\centering
\includegraphics[width=2.3in]{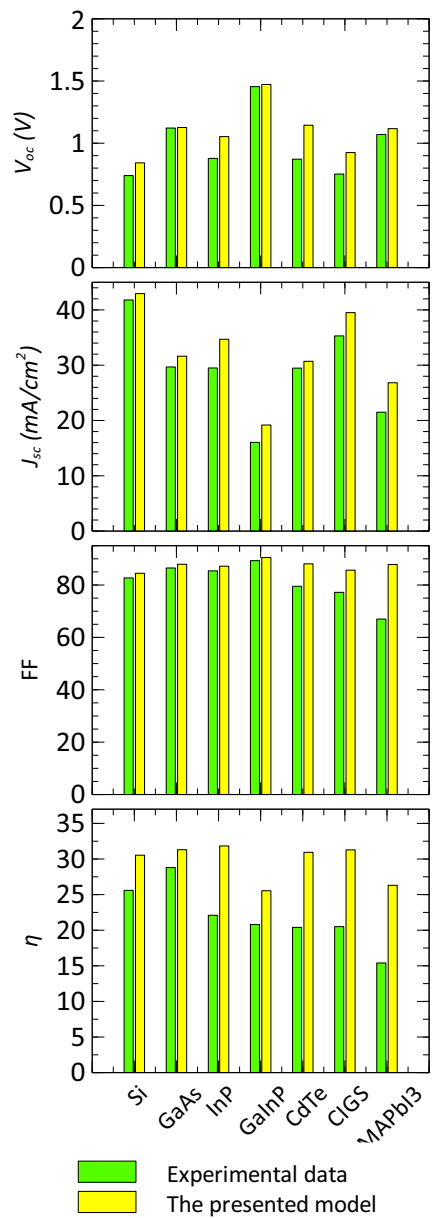}
\caption{Comparison for non-excitonic solar cell performances of the relevant reference materials as estimated by the presented and as experimental published for the best reported cells.}
\label{Comp3}
\end{figure}

In the next analysis, the improved model is applied to the reference non-excitonic materials. As aforementioned, the original Scharber model was designed for OPVs. It can be adjusted to also work for inorganic cells in a similar way as our proposed model, i.e. by working with two sets of parameters. Fig. \ref{Comp3} shows a comparison between the reported experimental efficiencies and the estimated ones by the proposed model for non-excitonic cells. Again, the model provides very good estimations. For the well optimized devices like Si and GaAs, the presented model suggests that the room of improvement is limited. However, it indicates that there is a possibility to considerably improve the performance of \ce{MAPbI3}, CIGS, CdTe, and InP.

\begin{figure}[hp]
\centering
\includegraphics[width=2.3in]{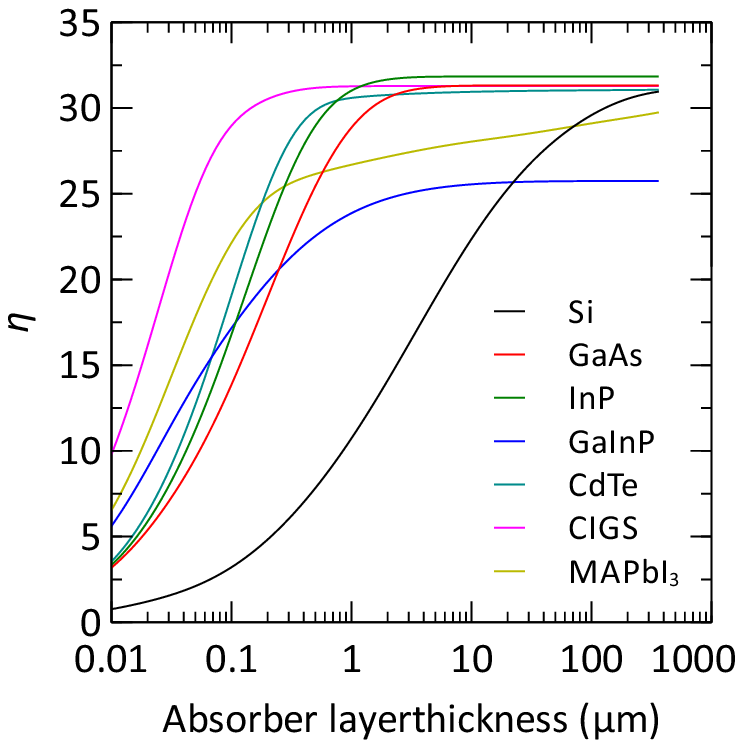}
\caption{The estimated non-excitonic solar cell efficiencies for the relevant reference materials as a function of the absorbing layer thickness.}
\label{Comp4}
\end{figure}

The last analysis is for the effect the absorber layer thickness on the expected efficiencies for the non-excitonic solar cell materials. The results are shown in Fig. \ref{Comp4}. The expected efficiencies for all the studied materials saturate after few $\mu$m except for Si solar cells. It takes very thick layer to reach a reasonable efficiency. As known, this is due to its weak absorption \cite{FH1,FH2,C03,M03}.

\section{Conclusion}

A descriptor model for solar cell efficiencies estimation is developed. Relative to the original Scharber model, the developed model presented here revisits the three main performance factors ($V_{oc}$, $J_{sc}$, and $FF$; Eq. \ref{SCGenEff}). For the short circuit current ($J_{sc}$), the model takes in full account the details of the absorption spectrum $\alpha(E$) to evaluate the photo-generated current ($J_g$), and uses new and more elaborate parametrization for the other components contributing to this quantity, i.e. the scattering distribution and the diffusion length, which characterize the transport and the recombination. The open current voltage $V_{oc}$ is expressed in terms of a power series of the energy gap fitted against available experimental data. The fill factor is estimated using adjusted empirical model originally suggested by M. Green \cite{G06}. Using two different sets of parameters, the model can be used for both, excitonic and non-excitonic materials.

The analysis of the new model shows that its much better performance arises from the improved predictions for the open circuit voltage and the short circuit current ($V_{oc}$ and $J_{sc}$). On the other hand, the estimation of the original Scharber model for the fill factor ($FF$) is  slightly better than our model when compared to the experimentally reported values. However, since $FF$ depends extremely on the device design and optimization, the larger values estimated by the proposed model indicate that the performance of the reference cells can be improved by proper device optimization.

We expect the proposed descriptor model to allow for more accurate assessments of the performance of light harvesting materials. Even though the results of material screening efforts based on this model are still missing, the model was already shown to be useful for the study of the response of the performance of a given material to changing some parameters of the device design such as layer thickness. 

\section*{Appendix A}
\label{appndx}

The following tables show the efficiencies and cell performances for some of the best reported solar cells. For some materials, better efficiencies were reported but without full details. So, we limit the analysis for the reported cell with full performance parameters.

\begin{table*} [hpt]
\small
\centering
\renewcommand{\arraystretch}{1.2}
\caption{The efficiencies and cell performances for the most-efficient reported  inorganic homo junction solar cells with full data.}
\begin{tabular}{|c|c|c|c|c|c|c|l|}
\hline 
• & $E_g$ (eV) & L ($\mu$ m) & $V_{oc}$ (V) & $J_{sc}$ (mA/cm$^2$) & FF & $\eta$ & Ref. \\ 
\hline 
\ce{Si} & 1.12 & 200 & 0.74 & 41.8 & 82.7 & 25.6 & \cite{M03} \\  
\ce{GaAs} & 1.43 & 1 & 1.122 & 29.68 & 86.5 & 28.8 & \cite{K05} \\ 
\ce{InP} & 1.35 & 3  & 0.878 & 29.5 & 85.4 & 22.1 & \cite{K06} \\ 
\hline 
\end{tabular} 
\label{InorgHmJ}
\end{table*}

\begin{table*} [hpt]
\small
\centering
\renewcommand{\arraystretch}{1.2}
\caption{The efficiencies and cell performances for the most-efficient reported inorganic and hybrid heterojunction junction solar cells with full data.}
\begin{tabular}{|c|c|c|c|c|c|c|l|}
\hline 
• & $E_g$ (eV) & L ($\mu$ m) & $V_{oc}$ (V) & $J_{sc}$ (mA/cm$^2$) & FF & $\eta$ & Ref. \\ 
\hline 
\ce{GaInP} & 1.81 & 1 	 & 1.455  & 16.04 & 89.3 & 20.8 & \cite{G07} \\ 
\ce{CdTe} & 1.45 & 3 	 & 0.872 & 29.47 & 79.5 & 20.4 & \cite{H04} \\ 
\ce{CIGS} & 1.21 & 2 	 & 0.752 & 35.3 & 77.2 & 20.5 & \cite{O02} \\ 
\ce{MAPbI3} & 1.42 & 0.3 & 1.07 & 21.5 & 67.0 & 15.4 & \cite{L02} \\ 
\hline
\end{tabular} 
\label{InrgHtJ}
\end{table*}

\begin{table*} [hpt]
\small
\centering
\renewcommand{\arraystretch}{1.2}
\caption{The efficiencies and cell performances for some of the most-efficient reported organic solar cells with full data.}
\begin{tabular}{|c|c|c|c|c|c|c|l|}
\hline 
• & $E_g$ (eV) & $V_{oc}$ (V) & $J_{sc}$ (mA/cm$^2$) & FF & $\eta$ & Ref. \\ 
\hline 
\ce{SQ} & 1.47 & 0.86 & 13.6 & 52 & 6.1 & \cite{C04} \\ 
\ce{DTS} & 1.40 & 0.78 & 14.4 & 59.3 & 6.7 & \cite{S06} \\ 
\ce{CuPc} & 1.46 & 0.54 & 15 & 61 & 5.4 & \cite{X01} \\ 
\ce{ZnPc} & 1.39 & 0.66 & 12 & 64 & 5.0 & \cite{F05} \\ 
\ce{DBP} & 1.63 & 1.93 & 13.2 & 66 & 8.1 & \cite{X02} \\ 
P3HT & 1.77 & 0.87 & 11.35 & 75 & 7.4 & \cite{G08} \\ 
PTB7 & 1.51 & 0.75 & 17.46 & 70 & 9.2 & \cite{H05} \\ 
\hline 
\end{tabular} 
\label{OrgHtJ}
\end{table*}

\bibliography{ImpSchModel}{}

\begin{thebibliography}{10}
\expandafter\ifx\csname url\endcsname\relax
  \def\url#1{\texttt{#1}}\fi
\expandafter\ifx\csname urlprefix\endcsname\relax\def\urlprefix{URL }\fi
\providecommand{\bibinfo}[2]{#2}
\providecommand{\eprint}[2][]{\url{#2}}

\bibitem{A02}
\bibinfo{author}{Azimi, H.}, \bibinfo{author}{Hou, Y.} \&
  \bibinfo{author}{Brabec, C.~J.}
\newblock \bibinfo{title}{Towards low-cost, environmentally friendly printed
  chalcopyrite and kesterite solar cells}.
\newblock \emph{\bibinfo{journal}{Energy Environ. Sci.}}
  \textbf{\bibinfo{volume}{7}}, \bibinfo{pages}{1829--1849}
  (\bibinfo{year}{2014}).

\bibitem{H01}
\bibinfo{author}{Habas, S.~E.}, \bibinfo{author}{Platt, H.~A.},
  \bibinfo{author}{van Hest, M.~F.} \& \bibinfo{author}{Ginley, D.~S.}
\newblock \bibinfo{title}{Low-cost inorganic solar cells: from ink to printed
  device}.
\newblock \emph{\bibinfo{journal}{Chem. Rev.}} \textbf{\bibinfo{volume}{110}},
  \bibinfo{pages}{6571--6594} (\bibinfo{year}{2010}).

\bibitem{W02}
\bibinfo{author}{Wadia, C.}, \bibinfo{author}{Alivisatos, A.~P.} \&
  \bibinfo{author}{Kammen, D.~M.}
\newblock \bibinfo{title}{Materials availability expands the opportunity for
  large-scale photovoltaics deployment}.
\newblock \emph{\bibinfo{journal}{Energy Environ. Sci.}}
  \textbf{\bibinfo{volume}{43}}, \bibinfo{pages}{2072--2077}
  (\bibinfo{year}{2009}).

\bibitem{FH1}
\bibinfo{author}{Alharbi, F.} \emph{et~al.}
\newblock \bibinfo{title}{Abundant non-toxic materials for thin film solar
  cells: Alternative to conventional materials}.
\newblock \emph{\bibinfo{journal}{Renewable Energy}}
  \textbf{\bibinfo{volume}{36}}, \bibinfo{pages}{2753--2758}
  (\bibinfo{year}{2011}).

\bibitem{FH2}
\bibinfo{author}{Hossain, M.} \& \bibinfo{author}{Alharbi, F.}
\newblock \bibinfo{title}{Recent advances in alternative material
  photovoltaics}.
\newblock \emph{\bibinfo{journal}{Mater. Technol.}}
  \textbf{\bibinfo{volume}{28}}, \bibinfo{pages}{88--97}
  (\bibinfo{year}{2013}).

\bibitem{D01}
\bibinfo{author}{Dimroth, F.} \emph{et~al.}
\newblock \bibinfo{title}{Wafer bonded four-junction gainp/gaas//gainasp/gainas
  concentrator solar cells with 44.7\% efficiency}.
\newblock \emph{\bibinfo{journal}{Prog. Photovoltaics Res. Appl.}}
  \textbf{\bibinfo{volume}{22}}, \bibinfo{pages}{277--282}
  (\bibinfo{year}{2014}).

\bibitem{S01}
\bibinfo{author}{Snaith, H.~J.}
\newblock \bibinfo{title}{Perovskites: the emergence of a new era for low-cost,
  high-efficiency solar cells}.
\newblock \emph{\bibinfo{journal}{J. Phys. Chem. Lett.}}
  \textbf{\bibinfo{volume}{4}}, \bibinfo{pages}{3623--3630}
  (\bibinfo{year}{2013}).

\bibitem{N01}
\bibinfo{author}{Nelson, J.}, \bibinfo{author}{Kirkpatrick, J.} \&
  \bibinfo{author}{Ravirajan, P.}
\newblock \bibinfo{title}{Factors limiting the efficiency of molecular
  photovoltaic devices}.
\newblock \emph{\bibinfo{journal}{Phys. Rev. B}} \textbf{\bibinfo{volume}{69}},
  \bibinfo{pages}{035337} (\bibinfo{year}{2004}).

\bibitem{M01}
\bibinfo{author}{Markvart, T.}
\newblock \bibinfo{title}{The thermodynamics of optical {\'e}tendue}.
\newblock \emph{\bibinfo{journal}{J. Opt. A}} \textbf{\bibinfo{volume}{10}},
  \bibinfo{pages}{015008} (\bibinfo{year}{2008}).

\bibitem{K01}
\bibinfo{author}{Kirk, A.}
\newblock \bibinfo{title}{A discussion of fundamental solar photovoltaic cell
  physics}.
\newblock \emph{\bibinfo{journal}{Physica B}} \textbf{\bibinfo{volume}{423}},
  \bibinfo{pages}{58--59} (\bibinfo{year}{2013}).

\bibitem{FH3}
\bibinfo{author}{Alharbi, F.~H.} \& \bibinfo{author}{Kais, S.}
\newblock \bibinfo{title}{Theoretical limits of photovoltaics efficiency and
  possible improvements by intuitive approaches learned from photosynthesis and
  quantum coherence}.
\newblock \emph{\bibinfo{journal}{Renewable Sustainable Energy Rev.}}
  \textbf{\bibinfo{volume}{43}}, \bibinfo{pages}{1073 -- 1089}
  (\bibinfo{year}{2015}).

\bibitem{H02}
\bibinfo{author}{Hachmann, J.} \emph{et~al.}
\newblock \bibinfo{title}{The harvard clean energy project: large-scale
  computational screening and design of organic photovoltaics on the world
  community grid}.
\newblock \emph{\bibinfo{journal}{J. Phys. Chem. Lett.}}
  \textbf{\bibinfo{volume}{2}}, \bibinfo{pages}{2241--2251}
  (\bibinfo{year}{2011}).

\bibitem{O01}
\bibinfo{author}{Olivares-Amaya, R.} \emph{et~al.}
\newblock \bibinfo{title}{Accelerated computational discovery of
  high-performance materials for organic photovoltaics by means of
  cheminformatics}.
\newblock \emph{\bibinfo{journal}{Energy Environ. Sci.}}
  \textbf{\bibinfo{volume}{4}}, \bibinfo{pages}{4849--4861}
  (\bibinfo{year}{2011}).

\bibitem{H07}
\bibinfo{author}{Hachmann, J.} \emph{et~al.}
\newblock \bibinfo{title}{Lead candidates for high-performance organic
  photovoltaics from high-throughput quantum chemistry--the harvard clean
  energy project}.
\newblock \emph{\bibinfo{journal}{Energy Environ. Sci.}}
  \textbf{\bibinfo{volume}{7}}, \bibinfo{pages}{698--704}
  (\bibinfo{year}{2014}).

\bibitem{L03}
\bibinfo{author}{Landis, D.~D.} \emph{et~al.}
\newblock \bibinfo{title}{The computational materials repository}.
\newblock \emph{\bibinfo{journal}{Comput. Sci. Eng.}}
  \textbf{\bibinfo{volume}{14}}, \bibinfo{pages}{51--57}
  (\bibinfo{year}{2012}).

\bibitem{C06}
\bibinfo{author}{Castelli, I.~E.} \emph{et~al.}
\newblock \bibinfo{title}{Computational screening of perovskite metal oxides
  for optimal solar light capture}.
\newblock \emph{\bibinfo{journal}{Energy Environ. Sci.}}
  \textbf{\bibinfo{volume}{5}}, \bibinfo{pages}{5814--5819}
  (\bibinfo{year}{2012}).

\bibitem{C07}
\bibinfo{author}{Curtarolo, S.} \emph{et~al.}
\newblock \bibinfo{title}{The high-throughput highway to computational
  materials design}.
\newblock \emph{\bibinfo{journal}{Nat. Mater.}} \textbf{\bibinfo{volume}{12}},
  \bibinfo{pages}{191--201} (\bibinfo{year}{2013}).

\bibitem{SM1}
\bibinfo{author}{Scharber, M.~C.} \emph{et~al.}
\newblock \bibinfo{title}{Design rules for donors in bulk-heterojunction solar
  cells—towards 10\% energy-conversion efficiency}.
\newblock \emph{\bibinfo{journal}{Adv. Mater.}} \textbf{\bibinfo{volume}{18}},
  \bibinfo{pages}{789--794} (\bibinfo{year}{2006}).

\bibitem{H03}
\bibinfo{author}{Hoppe, H.} \& \bibinfo{author}{Sariciftci, N.~S.}
\newblock \bibinfo{title}{Organic solar cells: An overview}.
\newblock \emph{\bibinfo{journal}{J. Mater. Res.}}
  \textbf{\bibinfo{volume}{19}}, \bibinfo{pages}{1924--1945}
  (\bibinfo{year}{2004}).

\bibitem{F03}
\bibinfo{author}{Forrest, S.~R.}
\newblock \bibinfo{title}{The limits to organic photovoltaic cell efficiency}.
\newblock \emph{\bibinfo{journal}{MRS Bull.}} \textbf{\bibinfo{volume}{30}},
  \bibinfo{pages}{28--32} (\bibinfo{year}{2005}).

\bibitem{V01}
\bibinfo{author}{Vajjala{\'a}Kesava, S.} \emph{et~al.}
\newblock \bibinfo{title}{Direct measurements of exciton diffusion length
  limitations on organic solar cell performance}.
\newblock \emph{\bibinfo{journal}{Chem. Commun.}}
  \textbf{\bibinfo{volume}{48}}, \bibinfo{pages}{5859--5861}
  (\bibinfo{year}{2012}).

\bibitem{A03}
\bibinfo{title}{{ASTM G173-03}: Standard tables for reference solar spectral
  irradiances: Direct normal and hemispherical on 37$^{\circ}$ tilted surface}.
\newblock \bibinfo{type}{Tech. Rep.} (\bibinfo{year}{2008}).
\newblock \bibinfo{note}{Http://www.astm.org/Standards/G173.htm}.

\bibitem{A01}
\bibinfo{author}{Abrams, Z.~R.}, \bibinfo{author}{Gharghi, M.},
  \bibinfo{author}{Niv, A.}, \bibinfo{author}{Gladden, C.} \&
  \bibinfo{author}{Zhang, X.}
\newblock \bibinfo{title}{Theoretical efficiency of 3$^{rd}$ generation solar
  cells: Comparison between carrier multiplication and down-conversion}.
\newblock \emph{\bibinfo{journal}{Sol. Energy Mater. Sol. Cells}}
  \textbf{\bibinfo{volume}{99}}, \bibinfo{pages}{308--315}
  (\bibinfo{year}{2012}).

\bibitem{W01}
\bibinfo{author}{W{\"u}rfel, P.}, \bibinfo{author}{Brown, A.},
  \bibinfo{author}{Humphrey, T.} \& \bibinfo{author}{Green, M.}
\newblock \bibinfo{title}{Particle conservation in the hot-carrier solar cell}.
\newblock \emph{\bibinfo{journal}{Prog. Photovoltaics Res. Appl.}}
  \textbf{\bibinfo{volume}{13}}, \bibinfo{pages}{277--285}
  (\bibinfo{year}{2005}).

\bibitem{B01}
\bibinfo{author}{Baruch, P.}, \bibinfo{author}{De~Vos, A.},
  \bibinfo{author}{Landsberg, P.} \& \bibinfo{author}{Parrott, J.}
\newblock \bibinfo{title}{On some thermodynamic aspects of photovoltaic solar
  energy conversion}.
\newblock \emph{\bibinfo{journal}{Sol. Energy Mater. Sol. Cells}}
  \textbf{\bibinfo{volume}{36}}, \bibinfo{pages}{201--222}
  (\bibinfo{year}{1995}).

\bibitem{F01}
\bibinfo{author}{Fan, J.}, \bibinfo{author}{Tsaur, B.} \&
  \bibinfo{author}{Palm, B.}
\newblock \bibinfo{title}{Optimum design of high efficiency tandem solar
  cells}.
\newblock In \emph{\bibinfo{booktitle}{Proc. 16th IEEE Photovoltaic Specialists
  Conference}}, \bibinfo{pages}{692--701} (\bibinfo{year}{1982}).

\bibitem{N02}
\bibinfo{author}{Nell, M.~E.} \& \bibinfo{author}{Barnett, A.~M.}
\newblock \bibinfo{title}{The spectral pn junction model for tandem solar-cell
  design}.
\newblock \emph{\bibinfo{journal}{IEEE Trans. Electron Devices}}
  \textbf{\bibinfo{volume}{34}}, \bibinfo{pages}{257--266}
  (\bibinfo{year}{1987}).

\bibitem{W03}
\bibinfo{author}{Wanlass, M.} \emph{et~al.}
\newblock \bibinfo{title}{Practical considerations in tandem cell modeling}.
\newblock \emph{\bibinfo{journal}{Sol. Cells}} \textbf{\bibinfo{volume}{27}},
  \bibinfo{pages}{191--204} (\bibinfo{year}{1989}).

\bibitem{C01}
\bibinfo{author}{Coutts, T.}
\newblock \bibinfo{title}{A review of progress in thermophotovoltaic generation
  of electricity}.
\newblock \emph{\bibinfo{journal}{Renewable Sustainable Energy Rev.}}
  \textbf{\bibinfo{volume}{3}}, \bibinfo{pages}{77--184}
  (\bibinfo{year}{1999}).

\bibitem{C05}
\bibinfo{author}{Coutts, T.~J.}, \bibinfo{author}{Emery, K.~A.} \&
  \bibinfo{author}{Scott~Ward, J.}
\newblock \bibinfo{title}{Modeled performance of polycrystalline thin-film
  tandem solar cells}.
\newblock \emph{\bibinfo{journal}{Prog. Photovoltaics Res. Appl.}}
  \textbf{\bibinfo{volume}{10}}, \bibinfo{pages}{195--203}
  (\bibinfo{year}{2002}).

\bibitem{P02}
\bibinfo{author}{Potscavage~Jr, W.~J.}, \bibinfo{author}{Yoo, S.} \&
  \bibinfo{author}{Kippelen, B.}
\newblock \bibinfo{title}{Origin of the open-circuit voltage in multilayer
  heterojunction organic solar cells}.
\newblock \emph{\bibinfo{journal}{Appl. Phys. Lett.}}
  \textbf{\bibinfo{volume}{93}}, \bibinfo{pages}{193308}
  (\bibinfo{year}{2008}).

\bibitem{K07}
\bibinfo{author}{Katz, E.} \emph{et~al.}
\newblock \bibinfo{title}{Temperature dependence for the photovoltaic device
  parameters of polymer-fullerene solar cells under operating conditions}.
\newblock \emph{\bibinfo{journal}{J. Appl. Phys.}}
  \textbf{\bibinfo{volume}{90}}, \bibinfo{pages}{5343--5350}
  (\bibinfo{year}{2001}).

\bibitem{S02}
\bibinfo{author}{Stutenbaeumer, U.} \& \bibinfo{author}{Lewetegn, E.}
\newblock \bibinfo{title}{Comparison of minority carrier diffusion length
  measurements in silicon solar cells by the photo-induced open-circuit voltage
  decay (ocvd) with different excitation sources}.
\newblock \emph{\bibinfo{journal}{Renewable Energy}}
  \textbf{\bibinfo{volume}{20}}, \bibinfo{pages}{65--74}
  (\bibinfo{year}{2000}).

\bibitem{F02}
\bibinfo{author}{Fortini, A.}, \bibinfo{author}{Lande, R.},
  \bibinfo{author}{Madelon, R.} \& \bibinfo{author}{Bauduin, P.}
\newblock \bibinfo{title}{Carrier concentration and diffusion length
  measurements by 8-mm-microwave magnetophotoreflectivity in germanium and
  silicon}.
\newblock \emph{\bibinfo{journal}{J. Appl. Phys.}}
  \textbf{\bibinfo{volume}{45}}, \bibinfo{pages}{3380--3384}
  (\bibinfo{year}{1974}).

\bibitem{L01}
\bibinfo{author}{Leon, R.}
\newblock \bibinfo{title}{Diffusion length measurement in bulk and epitaxially
  grown iii-v semiconductors using charge collection microscopy}.
\newblock In \emph{\bibinfo{booktitle}{Proc. 19th IEEE Photovoltaic Specialists
  Conference}}, vol.~\bibinfo{volume}{1}, \bibinfo{pages}{808--812}
  (\bibinfo{year}{1987}).

\bibitem{M02}
\bibinfo{author}{Matar{\'e}, H.} \& \bibinfo{author}{Wolff, G.}
\newblock \bibinfo{title}{Concentration enhancement of current density and
  diffusion length in iii--v ternary compound solar cells}.
\newblock \emph{\bibinfo{journal}{Appl. Phys. A}}
  \textbf{\bibinfo{volume}{17}}, \bibinfo{pages}{335--342}
  (\bibinfo{year}{1978}).

\bibitem{G01}
\bibinfo{author}{Gautron, J.} \& \bibinfo{author}{Lemasson, P.}
\newblock \bibinfo{title}{Photoelectrochemical determination of minority
  carrier diffusion length in ii--vi compounds}.
\newblock \emph{\bibinfo{journal}{J. Cryst. Growth}}
  \textbf{\bibinfo{volume}{59}}, \bibinfo{pages}{332--337}
  (\bibinfo{year}{1982}).

\bibitem{S03}
\bibinfo{author}{Stranks, S.~D.} \emph{et~al.}
\newblock \bibinfo{title}{Electron-hole diffusion lengths exceeding 1
  micrometer in an organometal trihalide perovskite absorber}.
\newblock \emph{\bibinfo{journal}{Science}} \textbf{\bibinfo{volume}{342}},
  \bibinfo{pages}{341--344} (\bibinfo{year}{2013}).

\bibitem{S04}
\bibinfo{author}{Sze, S.~M.} \& \bibinfo{author}{Ng, K.~K.}
\newblock \emph{\bibinfo{title}{Physics of semiconductor devices}}
  (\bibinfo{publisher}{John Wiley \& Sons}, \bibinfo{year}{2006}).

\bibitem{W04}
\bibinfo{author}{Weber, M.~J.}
\newblock \emph{\bibinfo{title}{Handbook of optical materials}},
  vol.~\bibinfo{volume}{19} (\bibinfo{publisher}{CRC press},
  \bibinfo{year}{2002}).

\bibitem{T01}
\bibinfo{author}{Theodoropoulou, S.}, \bibinfo{author}{Papadimitriou, D.},
  \bibinfo{author}{Anestou, K.}, \bibinfo{author}{Cobet, C.} \&
  \bibinfo{author}{Esser, N.}
\newblock \bibinfo{title}{Optical properties of cuin$_{1- x}$ga$_x$se2
  quaternary alloys for solar-energy conversion}.
\newblock \emph{\bibinfo{journal}{Semicond. Sci. Technol.}}
  \textbf{\bibinfo{volume}{24}}, \bibinfo{pages}{015014}
  (\bibinfo{year}{2009}).

\bibitem{G04}
\bibinfo{author}{Gamliel, S.} \& \bibinfo{author}{Etgar, L.}
\newblock \bibinfo{title}{Organo-metal perovskite based solar cells: sensitized
  versus planar architecture}.
\newblock \emph{\bibinfo{journal}{RSC Adv.}} \textbf{\bibinfo{volume}{4}},
  \bibinfo{pages}{29012--29021} (\bibinfo{year}{2014}).

\bibitem{W05}
\bibinfo{author}{Wei, G.} \emph{et~al.}
\newblock \bibinfo{title}{Functionalized squaraine donors for nanocrystalline
  organic photovoltaics}.
\newblock \emph{\bibinfo{journal}{ACS Nano}} \textbf{\bibinfo{volume}{6}},
  \bibinfo{pages}{972--978} (\bibinfo{year}{2011}).

\bibitem{D02}
\bibinfo{author}{Duch{\'e}, D.} \emph{et~al.}
\newblock \bibinfo{title}{Optical performance and color investigations of
  hybrid solar cells based on p3ht: Zno, pcpdtbt: Zno, ptb7: Zno and dts
  (ptth$_2$)$_2$: Zno}.
\newblock \emph{\bibinfo{journal}{Sol. Energy Mater. Sol. Cells}}
  \textbf{\bibinfo{volume}{126}}, \bibinfo{pages}{197--204}
  (\bibinfo{year}{2014}).

\bibitem{F04}
\bibinfo{author}{Fujishima, D.} \emph{et~al.}
\newblock \bibinfo{title}{Organic thin-film solar cell employing a novel
  electron-donor material}.
\newblock \emph{\bibinfo{journal}{Sol. Energy Mater. Sol. Cells}}
  \textbf{\bibinfo{volume}{93}}, \bibinfo{pages}{1029--1032}
  (\bibinfo{year}{2009}).

\bibitem{S05}
\bibinfo{author}{Senthilarasu, S.} \emph{et~al.}
\newblock \bibinfo{title}{Characterization of zinc phthalocyanine (znpc) for
  photovoltaic applications}.
\newblock \emph{\bibinfo{journal}{Appl. Phys. A}}
  \textbf{\bibinfo{volume}{77}}, \bibinfo{pages}{383--389}
  (\bibinfo{year}{2003}).

\bibitem{G02}
\bibinfo{author}{Gregg, B.~A.}
\newblock \bibinfo{title}{Excitonic solar cells}.
\newblock \emph{\bibinfo{journal}{J. Phys. Chem. B}}
  \textbf{\bibinfo{volume}{107}}, \bibinfo{pages}{4688--4698}
  (\bibinfo{year}{2003}).

\bibitem{G03}
\bibinfo{author}{Gregg, B.~A.}
\newblock \bibinfo{title}{The photoconversion mechanism of excitonic solar
  cells}.
\newblock \emph{\bibinfo{journal}{MRS Bull.}} \textbf{\bibinfo{volume}{30}},
  \bibinfo{pages}{20--22} (\bibinfo{year}{2005}).

\bibitem{P03}
\bibinfo{author}{Poruba, A.} \emph{et~al.}
\newblock \bibinfo{title}{Optical absorption and light scattering in
  microcrystalline silicon thin films and solar cells}.
\newblock \emph{\bibinfo{journal}{J. Appl. Phys.}}
  \textbf{\bibinfo{volume}{88}}, \bibinfo{pages}{148--160}
  (\bibinfo{year}{2000}).

\bibitem{K02}
\bibinfo{author}{Krc, J.}, \bibinfo{author}{Zeman, M.}, \bibinfo{author}{Smole,
  F.} \emph{et~al.}
\newblock \bibinfo{title}{Optical modelling of thin-film silicon solar cells
  deposited on textured substrates}.
\newblock \emph{\bibinfo{journal}{Thin Solid Films}}
  \textbf{\bibinfo{volume}{451}}, \bibinfo{pages}{298--302}
  (\bibinfo{year}{2004}).

\bibitem{K03}
\bibinfo{author}{Krc, J.}, \bibinfo{author}{Smole, F.} \emph{et~al.}
\newblock \bibinfo{title}{Analysis of light scattering in amorphous si: H solar
  cells by a one-dimensional semi-coherent optical model}.
\newblock \emph{\bibinfo{journal}{Prog. Photovoltaics Res. Appl.}}
  \textbf{\bibinfo{volume}{11}}, \bibinfo{pages}{15--26}
  (\bibinfo{year}{2003}).

\bibitem{M04}
\bibinfo{author}{M{\"u}ller, J.}, \bibinfo{author}{Rech, B.},
  \bibinfo{author}{Springer, J.} \& \bibinfo{author}{Vanecek, M.}
\newblock \bibinfo{title}{Tco and light trapping in silicon thin film solar
  cells}.
\newblock \emph{\bibinfo{journal}{Sol. Energy}} \textbf{\bibinfo{volume}{77}},
  \bibinfo{pages}{917--930} (\bibinfo{year}{2004}).

\bibitem{B02}
\bibinfo{author}{Battaglia, C.} \emph{et~al.}
\newblock \bibinfo{title}{Light trapping in solar cells: can periodic beat
  random?}
\newblock \emph{\bibinfo{journal}{ACS Nano}} \textbf{\bibinfo{volume}{6}},
  \bibinfo{pages}{2790--2797} (\bibinfo{year}{2012}).

\bibitem{C02}
\bibinfo{author}{Callahan, D.~M.}, \bibinfo{author}{Munday, J.~N.} \&
  \bibinfo{author}{Atwater, H.~A.}
\newblock \bibinfo{title}{Solar cell light trapping beyond the ray optic
  limit}.
\newblock \emph{\bibinfo{journal}{Nano Lett.}} \textbf{\bibinfo{volume}{12}},
  \bibinfo{pages}{214--218} (\bibinfo{year}{2012}).

\bibitem{C03}
\bibinfo{author}{Cho, J.-S.} \emph{et~al.}
\newblock \bibinfo{title}{Effect of nanotextured back reflectors on light
  trapping in flexible silicon thin-film solar cells}.
\newblock \emph{\bibinfo{journal}{Sol. Energy Mater. Sol. Cells}}
  \textbf{\bibinfo{volume}{102}}, \bibinfo{pages}{50--57}
  (\bibinfo{year}{2012}).

\bibitem{K04}
\bibinfo{author}{Kowalczewski, P.}, \bibinfo{author}{Liscidini, M.} \&
  \bibinfo{author}{Andreani, L.~C.}
\newblock \bibinfo{title}{Light trapping in thin-film solar cells with randomly
  rough and hybrid textures}.
\newblock \emph{\bibinfo{journal}{Opt. Express}} \textbf{\bibinfo{volume}{21}},
  \bibinfo{pages}{A808--A820} (\bibinfo{year}{2013}).

\bibitem{H05}
\bibinfo{author}{He, Z.} \emph{et~al.}
\newblock \bibinfo{title}{Enhanced power-conversion efficiency in polymer solar
  cells using an inverted device structure}.
\newblock \emph{\bibinfo{journal}{Nat. Photonics}}
  \textbf{\bibinfo{volume}{6}}, \bibinfo{pages}{591--595}
  (\bibinfo{year}{2012}).

\bibitem{Q01}
\bibinfo{author}{Qi, B.} \& \bibinfo{author}{Wang, J.}
\newblock \bibinfo{title}{Fill factor in organic solar cells}.
\newblock \emph{\bibinfo{journal}{Phys. Chem. Chem. Phys.}}
  \textbf{\bibinfo{volume}{15}}, \bibinfo{pages}{8972--8982}
  (\bibinfo{year}{2013}).

\bibitem{T02}
\bibinfo{author}{Taretto, K.}, \bibinfo{author}{Soldera, M.} \&
  \bibinfo{author}{Troviano, M.}
\newblock \bibinfo{title}{Accurate explicit equations for the fill factor of
  real solar cells—applications to thin-film solar cells}.
\newblock \emph{\bibinfo{journal}{Prog. Photovoltaics Res. Appl.}}
  \textbf{\bibinfo{volume}{21}}, \bibinfo{pages}{1489--1498}
  (\bibinfo{year}{2013}).

\bibitem{G05}
\bibinfo{author}{Gupta, D.}, \bibinfo{author}{Mukhopadhyay, S.} \&
  \bibinfo{author}{Narayan, K.}
\newblock \bibinfo{title}{Fill factor in organic solar cells}.
\newblock \emph{\bibinfo{journal}{Sol. Energy Mater. Sol. Cells}}
  \textbf{\bibinfo{volume}{94}}, \bibinfo{pages}{1309--1313}
  (\bibinfo{year}{2010}).

\bibitem{G06}
\bibinfo{author}{Green, M.~A.}
\newblock \bibinfo{title}{Accuracy of analytical expressions for solar cell
  fill factors}.
\newblock \emph{\bibinfo{journal}{Sol. Cells}} \textbf{\bibinfo{volume}{7}},
  \bibinfo{pages}{337--340} (\bibinfo{year}{1982}).

\bibitem{D03}
\bibinfo{author}{Dou, L.} \emph{et~al.}
\newblock \bibinfo{title}{25$^{th}$ anniversary article: A decade of
  organic/polymeric photovoltaic research}.
\newblock \emph{\bibinfo{journal}{Adv. Mater.}} \textbf{\bibinfo{volume}{25}},
  \bibinfo{pages}{6642--6671} (\bibinfo{year}{2013}).

\bibitem{M03}
\bibinfo{author}{Masuko, K.} \emph{et~al.}
\newblock \bibinfo{title}{Achievement of more than 25\% conversion efficiency
  with crystalline silicon heterojunction solar cell}.
\newblock \emph{\bibinfo{journal}{IEEE J. Photovoltaics}}
  \textbf{\bibinfo{volume}{4}}, \bibinfo{pages}{1433--1435}
  (\bibinfo{year}{2014}).

\bibitem{K05}
\bibinfo{author}{Kayes, B.~M.} \emph{et~al.}
\newblock \bibinfo{title}{27.6\% conversion efficiency, a new record for
  single-junction solar cells under 1 sun illumination}.
\newblock In \emph{\bibinfo{booktitle}{Proc. 37th IEEE Photovoltaic Specialists
  Conference}}, \bibinfo{pages}{4--8} (\bibinfo{organization}{IEEE},
  \bibinfo{year}{2011}).

\bibitem{K06}
\bibinfo{author}{Keavney, C.}, \bibinfo{author}{Haven, V.} \&
  \bibinfo{author}{Vernon, S.}
\newblock \bibinfo{title}{Emitter structures in mocvd inp solar cells}.
\newblock In \emph{\bibinfo{booktitle}{Proc. 21st IEEE Photovoltaic Specialists
  Conference}}, \bibinfo{pages}{141--144} (\bibinfo{organization}{IEEE},
  \bibinfo{year}{1990}).

\bibitem{G07}
\bibinfo{author}{Geisz, J.}, \bibinfo{author}{Steiner, M.},
  \bibinfo{author}{Garc{\'\i}a, I.}, \bibinfo{author}{Kurtz, S.} \&
  \bibinfo{author}{Friedman, D.}
\newblock \bibinfo{title}{Enhanced external radiative efficiency for 20.8\%
  efficient single-junction gainp solar cells}.
\newblock \emph{\bibinfo{journal}{Appl. Phys. Lett.}}
  \textbf{\bibinfo{volume}{103}}, \bibinfo{pages}{041118}
  (\bibinfo{year}{2013}).

\bibitem{H04}
\bibinfo{author}{Huber, W.~H.} \& \bibinfo{author}{Duggal, A.~R.}
\newblock \bibinfo{title}{The renaissance of cdte-based photovoltaics}.
\newblock In \emph{\bibinfo{booktitle}{Thin Films for Solar and Energy
  Technology}}, vol. \bibinfo{volume}{9177} of \emph{\bibinfo{series}{SPIE
  Proc.}}

\bibitem{O02}
\bibinfo{author}{Osborne, M.}
\newblock \bibinfo{title}{Hanergy’s solibro has 20.5\% cigs solar cell
  verified by nrel}.
\newblock
  \bibinfo{howpublished}{\url{http://www.pv-tech.org/news/hanergys_solibro_has_20.5_cigs_solar_cell_verified_by_NREL}}
  (\bibinfo{year}{2014}).

\bibitem{L02}
\bibinfo{author}{Liu, M.}, \bibinfo{author}{Johnston, M.~B.} \&
  \bibinfo{author}{Snaith, H.~J.}
\newblock \bibinfo{title}{Efficient planar heterojunction perovskite solar
  cells by vapour deposition}.
\newblock \emph{\bibinfo{journal}{Nat.}} \textbf{\bibinfo{volume}{501}},
  \bibinfo{pages}{395--398} (\bibinfo{year}{2013}).

\bibitem{C04}
\bibinfo{author}{Chen, G.} \emph{et~al.}
\newblock \bibinfo{title}{Co-evaporated bulk heterojunction solar cells with>
  6.0\% efficiency}.
\newblock \emph{\bibinfo{journal}{Adv. Mater.}} \textbf{\bibinfo{volume}{24}},
  \bibinfo{pages}{2768--2773} (\bibinfo{year}{2012}).

\bibitem{S06}
\bibinfo{author}{Sun, Y.} \emph{et~al.}
\newblock \bibinfo{title}{Solution-processed small-molecule solar cells with
  6.7\% efficiency}.
\newblock \emph{\bibinfo{journal}{Nat. Mater.}} \textbf{\bibinfo{volume}{11}},
  \bibinfo{pages}{44--48} (\bibinfo{year}{2012}).

\bibitem{X01}
\bibinfo{author}{Xue, J.}, \bibinfo{author}{Rand, B.~P.},
  \bibinfo{author}{Uchida, S.} \& \bibinfo{author}{Forrest, S.~R.}
\newblock \bibinfo{title}{A hybrid planar--mixed molecular heterojunction
  photovoltaic cell}.
\newblock \emph{\bibinfo{journal}{Adv. Mater.}} \textbf{\bibinfo{volume}{17}},
  \bibinfo{pages}{66--71} (\bibinfo{year}{2005}).

\bibitem{F05}
\bibinfo{author}{Fleetham, T.~B.} \emph{et~al.}
\newblock \bibinfo{title}{Efficient zinc phthalocyanine/c60 heterojunction
  photovoltaic devices employing tetracene anode interfacial layers}.
\newblock \emph{\bibinfo{journal}{ACS Appl. Mater. Interfaces}}
  \textbf{\bibinfo{volume}{6}}, \bibinfo{pages}{7254--7259}
  (\bibinfo{year}{2014}).

\bibitem{X02}
\bibinfo{author}{Xiao, X.}, \bibinfo{author}{Bergemann, K.~J.},
  \bibinfo{author}{Zimmerman, J.~D.}, \bibinfo{author}{Lee, K.} \&
  \bibinfo{author}{Forrest, S.~R.}
\newblock \bibinfo{title}{Small-molecule planar-mixed heterojunction
  photovoltaic cells with fullerene-based electron filtering buffers}.
\newblock \emph{\bibinfo{journal}{Adv. Energy Mater.}}
  \textbf{\bibinfo{volume}{4}}, \bibinfo{pages}{1301557}
  (\bibinfo{year}{2014}).

\bibitem{G08}
\bibinfo{author}{Guo, X.} \emph{et~al.}
\newblock \bibinfo{title}{High efficiency polymer solar cells based on poly
  (3-hexylthiophene)/indene-c 70 bisadduct with solvent additive}.
\newblock \emph{\bibinfo{journal}{Energy Environ. Sci.}}
  \textbf{\bibinfo{volume}{5}}, \bibinfo{pages}{7943--7949}
  (\bibinfo{year}{2012}).

\end{thebibliography}
\bibliographystyle{naturemag}

\end{document}